\documentclass[journal,twocolumn]{IEEEtran}%
\pdfoutput=1
\usepackage{amsfonts}
\usepackage{amsmath}
\usepackage{amssymb}
\usepackage{graphicx}%
\providecommand{\U}[1]{\protect\rule{.1in}{.1in}}

\begin{document}

\title{Examples of minimal-memory, non-catastrophic quantum convolutional encoders}
\author{Mark M. Wilde,~Monireh Houshmand, and Saied Hosseini-Khayat\thanks{Mark
M.\ Wilde is with the School of Computer Science, McGill University, Montreal,
Quebec, Canada, H3A 2A7. Monireh Houshmand and Saied Hosseini-Khayat are with
the Department of Electrical Engineering, Ferdowsi University of Mashhad, Mashhad, Iran.
(E-mail: mark.wilde@mcgill.ca; monireh.houshmand@gmail.com;
shk@ieee.org)}}
\maketitle

\begin{abstract}
One of the most important open questions in the theory of quantum
convolutional coding is to determine a minimal-memory, non-catastrophic,
polynomial-depth convolutional encoder for an arbitrary quantum convolutional
code. Here, we present a technique that finds quantum convolutional encoders
with such desirable properties for several example quantum convolutional codes
(an exposition of our technique in full generality appears elsewhere). We
first show how to encode the well-studied Forney-Grassl-Guha (FGG) code with
an encoder that exploits just one memory qubit (the former Grassl-R\"{o}tteler
encoder requires 15 memory qubits). We then show how our technique can find an
online decoder corresponding to this encoder, and we also detail the operation
of our technique on a different example of a quantum convolutional code.
Finally, the reduction in memory for the FGG\ encoder makes it feasible to
simulate the performance of a quantum turbo code employing it, and we present
the results of such simulations.

\end{abstract}

\section{Introduction}

Quantum convolutional coding is a method for quantum error correction that
protects a stream of quantum information from the negative effects of
decoherence \cite{PhysRevLett.91.177902,isit2005forney,ieee2007forney}.
This approach is highly beneficial in a
quantum communication paradigm where the only decoherence is due to a noisy
quantum channel connecting a sender to a receiver. Since the original work on
quantum convolutional coding, researchers have contributed a notable
literature on several aspects of these codes:\ methods for encoding them
\cite{GR06b,GR06,GR07}, algebraic constructions of them
\cite{cwit2007aly,arx2007aly}, and variations that include resources such as
gauge qubits, entanglement shared between sender and receiver \cite{WB07,WB08}%
, and hybrid constructions encoding both classical and quantum
bits~\cite{WB08IEEE}. Further progress has led to a successful
\textquotedblleft quantization\textquotedblright\ of the classical turbo
coding theory~\cite{BGT93,BM96}---quantum serial turbo codes employing
constituent quantum convolutional codes appear to be capacity-approaching
codes in the standard \cite{PTO09} and entanglement-assisted
settings~\cite{WH10b}.

Despite this progress, an important question concerning the practical
implementation of a quantum convolutional code has remained
unanswered:\ \textit{Is there a way to
implement a given quantum convolutional code with a
non-catastrophic, efficient, minimal-memory encoder?} This question is not only important for reducing the
overhead needed to implement these codes with a coherent quantum device, but
it is also important for the algorithm used to decode them---the
post-processing time for the decoding algorithm is exponential in the number
of memory qubits (specifically, it is $O\left(  4^{m}N\right)  $ where $m$ is
the number of memory qubits and $N$ is the length of the code~\cite{PTO09}).
Any reduction in the size of the memory could become especially important in
the context of fault-tolerant quantum computing (if these codes are ever
exploited for this purpose) because decoding delays can translate into the
accumulation of more errors.

Past efforts have not given satisfactory answers to the aforementioned
question. The Ollivier-Tillich \cite{PhysRevLett.91.177902,arxiv2004olliv} and
Grassl-R\"{o}tteler \cite{GR06b,GR06,GR07}\ algorithms for encoding quantum
convolutional codes are useful methods for accomplishing their intended goals,
but they both have no concern for quantum memory consumption. Also, the
Grassl-R\"{o}tteler algorithm in general can potentially lead to an encoding
circuit with exponential depth~\cite{GR06b}. We have attempted to answer the
minimal-memory question in prior work \cite{HHW10,HH10}, but our former
approaches are suboptimal in general---these approaches begin with a
pearl-necklace encoder resulting from the Grassl-R\"{o}tteler algorithm, and
they then find a particular convolutional encoder that uses the minimal amount
of memory for that particular pearl-necklace encoder (thus, they are
suboptimal because the original Grassl-R\"{o}tteler pearl-necklace encoder
might not result in a minimal-memory encoder).

In this paper, we present a technique for finding a minimal-memory,
non-catastrophic quantum convolutional encoder for several examples of quantum
convolutional codes (a presentation of our technique in full generality
appears elsewhere~\cite{HHW10a}). Our first example encoder is for the
Forney-Grassl-Guha\ (FGG)\ code~\cite{isit2005forney,ieee2007forney}, and our
technique gives a dramatic memory-consumption reduction from 15 memory qubits
for the original Grassl-R\"{o}tteler encoding \cite{GR06b,W09}\ to
\textit{just one memory qubit}. Our encoding technique simultaneously
accomplishes three goals:\ it finds a minimal-memory encoder for a quantum
convolutional code, it can force the resulting encoder to be
non-catastrophic~\cite{PTO09,WH10b}, and it provides an efficient encoder in
the sense that its depth is only $O\left(  n^{2}\right)  $ where $n$ is the
number of qubits in a frame of the code. Interestingly, the formula for the
minimal number of memory qubits~\cite{HHW10a}\ bears a close relationship to a
formula that gives the minimal number of ebits needed to encode an
entanglement-assisted quantum code~\cite{arx2008wildeOEA,PhysRevA.79.062322}.
This connection to entanglement-assisted quantum coding is perhaps not
surprising because the memory qubits of the encoder are entangled for some
time with the qubits sent over the channel. We also show how to determine an
online decoder corresponding to the encoder for the FGG\ code. The technique
for finding it is in the same spirit as the technique for finding the online
encoder, but the online decoder takes special care to decode both the logical
operators and the stabilizer operators of the code. The dramatic reduction in
memory consumption for the FGG\ encoder makes it reasonable to simulate the
performance of a quantum turbo code employing this encoder for its constituent
codes, and we plot the results of simulations in this paper.

We structure this paper as follows. The next section details how to find a
minimal-memory, non-catastrophic encoder for the FGG\ code.
Section~\ref{sec:FGG-decoder} details how to find an online decoder
corresponding to the choice of encoder in Section~\ref{sec:FGG-encoder}. We
then detail how our technique finds a minimal-memory, non-catastrophic encoder
of a more complicated quantum convolutional code from Ref.~\cite{GR07}. The
final section discusses and plots the results of simulating the performance of
the rate-$1/9$ FGG\ quantum turbo code.

\section{A Minimal-Memory, Non-Catastrophic Encoder for the Forney-Grassl-Guha
Code}

\label{sec:FGG-encoder}%
\begin{figure*}
[ptb]
\begin{center}
\includegraphics[
width=7.2in
]
{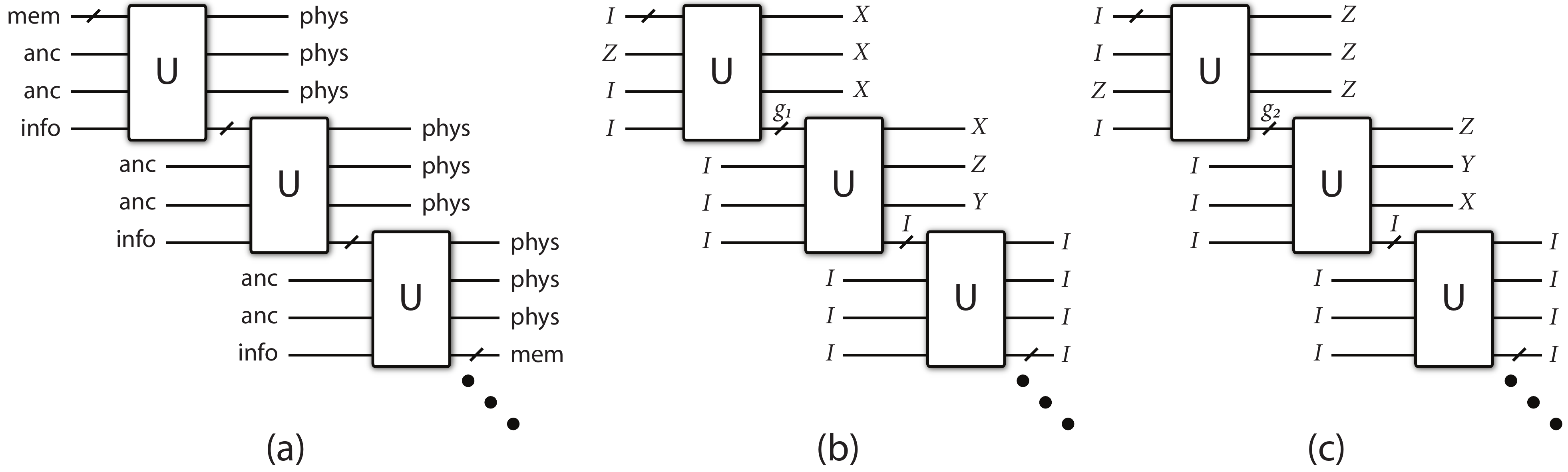}
\caption{The above figure provides a graphical aid
for understanding our technique that encodes the
Forney-Grassl-Guha (FGG)\ code with a minimal-memory encoder. (a) The
convolutional encoder $U$ for the FGG\ code should act on some unknown number $m$
of memory qubits labeled by \textquotedblleft mem\textquotedblright
\ (the diagonal slash through
a horizontal line indicates $m$ qubits), two
ancillas labeled by \textquotedblleft anc,\textquotedblright\ and one
information qubit labeled by \textquotedblleft info.\textquotedblright\ It
produces three output physical qubits labeled by \textquotedblleft
phys\textquotedblright\ and $m$ output memory qubits to be fed into the next
round of encoding. (b) The repeated application of the convolutional encoder
$U$ should transform the \textquotedblleft unencoded\textquotedblright
\ Pauli $Z$
operator acting on the first ancilla qubit to the first stabilizer generator
in (\ref{eq:FGG-stabilizer}). In order to so, the first application of the
encoder $U$ results in an intermediate, unspecified Pauli operator $g_{1}$
acting on the $m$\ output memory qubits. The second application of the encoder
$U$ completes the transformation of the unencoded input Pauli operator to the
first generator in (\ref{eq:FGG-stabilizer}). (c) The convolutional encoder
should similarly transform an \textquotedblleft unencoded\textquotedblright
\ $Z$ Pauli operator acting on the second ancilla qubit to the second
generator in (\ref{eq:FGG-stabilizer}). The shift-invariance of the encoder
guarantees that shifts of the unencoded $Z$ Pauli operators transform to
appropriate shifts of the generators in (\ref{eq:FGG-stabilizer}).}
\label{fig:fgg-technique}
\end{center}
\end{figure*}%
We first describe how our technique finds a minimal-memory, non-catastrophic
encoder for the FGG quantum convolutional
code~\cite{isit2005forney,ieee2007forney}. The stabilizer generators for this
quantum convolutional code are as follows:%
\begin{equation}%
\begin{array}
[c]{ccc}%
X & X & X\\
Z & Z & Z
\end{array}
\left\vert
\begin{array}
[c]{ccc}%
X & Z & Y\\
Z & Y & X
\end{array}
\right\vert \left.
\begin{array}
[c]{ccc}%
I & I & I\\
I & I & I
\end{array}
\right\vert \cdots,\label{eq:FGG-stabilizer}%
\end{equation}
where the vertical bars indicate that three qubits are in each frame and we
obtain the other generators of the code by shifting the above generators to
the right by multiples of three qubits. This code features two generators for
every three physical qubits, and so we should be able to encode it with a
quantum convolutional encoder of the form in Figure~\ref{fig:fgg-technique}%
(a). In particular, the encoder should act on some number $m$ of memory
qubits, two ancilla qubits, and one information qubit to produce three output
physical qubits and $m$ output memory qubits to be fed into the next round of encoding.

After inspecting Figure~\ref{fig:fgg-technique} and its caption, it should be
clear that an online encoder for the FGG code should transform the below Pauli
operators on the LHS to the ones on the RHS:%
\begin{equation}%
\begin{array}
[c]{cccc}%
I^{\otimes m} & Z & I & I\\
I^{\otimes m} & I & Z & I\\
g_{1} & I & I & I\\
g_{2} & I & I & I
\end{array}
\rightarrow%
\begin{array}
[c]{cccc}%
X & X & X & g_{1}\\
Z & Z & Z & g_{2}\\
X & Z & Y & I^{\otimes m}\\
Z & Y & X & I^{\otimes m}%
\end{array}
,\label{eq:FGG-transformation}%
\end{equation}
where $m$ is some unspecified number of memory qubits and $g_{1}$ and $g_{2}$
are $m$-qubit Pauli operators that we soon determine. The key observation to
make at this point is that any choice of the Clifford encoder $U$ should
preserve commutation relations. That is, if two $\left(  n+m\right)  $-fold
Pauli operators on the LHS commute, then the transformed versions of them on
the RHS should commute as well, and similarly if the input Pauli operators
in\ (\ref{eq:FGG-transformation}) anticommute. So, observe that the first two
input Pauli operators commute. This implies that the transformed versions of
them should commute, and $g_{1}$ and $g_{2}$ should thus anticommute to make
$XXXg_{1}$ and $ZZZg_{2}$ commute. Also, consider that the first and second
LHS rows in\ (\ref{eq:FGG-transformation}) commute with both the third and
fourth LHS rows. The corresponding output rows commute because the FGG\ code
is a valid quantum convolutional code. Finally, observe that the third and
fourth output rows anticommute. Thus, the third and fourth input rows should
anticommute, and this is consistent with our above observation that $g_{1}$
and $g_{2}$ should anticommute.

A sufficient choice for $g_{1}$ and $g_{2}$ to satisfy the above commutation
constraints is $g_{1}=X$ and$\ g_{2}=Z$, and this choice implies that the
Clifford encoder acts on just one memory qubit. This choice is optimal because
the encoder needs at least one memory qubit to encode the FGG\ code. Once we
have specified the Pauli operators $g_{1}$ and $g_{2}$, we can always find a
Clifford encoder that performs the transformation in
(\ref{eq:FGG-transformation}) because the Clifford group acts transitively on
Pauli operators (see Lemma~4 in Appendix~B\ of Ref.~\cite{BFG06}\ for an
explicit proof). A particular encoder that performs the transformation in
(\ref{eq:FGG-transformation}) is as follows:%
\begin{align}
&  \text{H}\left(  2\right)  \ \text{CNOT}\left(  4,1\right)  \ \text{H}%
\left(  4\right)  \ \text{CNOT}\left(  4,1\right)  \ \text{CNOT}\left(
4,2\right)  \nonumber\\
&  \ \text{CNOT}\left(  1,4\right)  \ \text{H}\left(  4\right)  \ \text{CNOT}%
\left(  3,4\right)  \ \text{P}\left(  1\right)  \ \text{CNOT}\left(
4,3\right)  \nonumber\\
&  \text{CNOT}\left(  1,3\right)  \ \text{CNOT}\left(  2,1\right)
\ \text{CNOT}\left(  2,3\right)  \ \text{CNOT}\left(  2,4\right)
,\label{eq:FGG-Clifford-encoder-gates}%
\end{align}
where the ordering of gates is from left to right and top to bottom,
$\text{H}(i)$ indicates a Hadamard gate acting on the $i^{\text{th}}$ qubit,
$\text{CNOT}\left(  i,j\right)  $ indicates a CNOT gate from the
$i^{\text{th}}$ qubit to the $j^{\text{th}}$ qubit, and $\text{P}(i)$
indicates a phase gate acting on the $i^{\text{th}}$ qubit. We found this
encoder by exploiting Grassl's algorithm for determining Clifford
unitaries~\cite{G06}.

We furthermore claim that any Clifford encoder performing the transformation
in (\ref{eq:FGG-transformation}) is non-catastrophic. Recall from
Refs.~\cite{PTO09,WH10b}\ that a catastrophic encoder is one whose state
diagram contains a cycle of zero physical weight that has non-zero logical
weight (this definition is in fact \textit{the same} as in the classical
case~\cite{V71}). Suppose for a contradiction that an encoder performing the
transformation is catastrophic. This would imply that the encoder creates some
cycle through memory states $h_{1}$, \ldots, $h_{p}$ of the following form:%
\[%
\begin{array}
[c]{cccc}%
h_{1} & s_{1,1} & s_{1,2} & l_{1}\\
\vdots & \vdots & \vdots & \vdots\\
h_{p} & s_{p,1} & s_{p,2} & l_{p}%
\end{array}
\rightarrow%
\begin{array}
[c]{cccc}%
I & I & I & h_{2}\\
\vdots & \vdots & \vdots & \vdots\\
I & I & I & h_{1}%
\end{array}
,
\]
where $h_{1}$, \ldots, $h_{p}$ can be arbitrary one-qubit Pauli operators
(since the memory consists of just one qubit), $s_{i,j}$ are one-qubit Pauli
operators equal to either the identity or Pauli $Z$, and $l_{i}$ are arbitrary
one-qubit Pauli operators with at least one of them not equal to the identity
operator. Observe that all of the output rows on the RHS above commute with
the last two rows on the RHS of the transformation in
(\ref{eq:FGG-transformation}). This observation implies that all of the rows
on the LHS above should commute with the last two rows on the LHS\ of the
transformation in (\ref{eq:FGG-transformation}). But this is only possible if
$h_{1}$, \ldots, $h_{p}$ are all equal to the one-qubit identity operator
because $g_{1}=X$ and $g_{2}=Z$. So all of the above entries are really just
cycles of the form $%
\begin{array}
[c]{cccc}%
I & s_{1} & s_{2} & l
\end{array}
\rightarrow%
\begin{array}
[c]{cccc}%
I & I & I & I
\end{array}
$. This input-output relation restricts $s_{1}$, $s_{2}$, and $l$ further---it
is impossible for $s_{1}$, $s_{2}$, and $l$ to be any Pauli operator besides
the identity operator. Otherwise, the encoder would not transform the entry on
the LHS to the all identity operator. Thus, the only cycle of zero-physical
weight in an encoder that implements the transformation in
(\ref{eq:FGG-transformation}) is the self-loop at the identity memory state
with zero logical weight, contradicting our original assumption that the
encoder is catastrophic.

\section{Online Decoder for the FGG Code}

\label{sec:FGG-decoder}We could simply use the inverse of the convolutional
encoder as the decoding unitary for the FGG code. Quantum turbo codes exploit
such a structure for the decoding (the decoding unitary there is actually the
inverse of the first convolutional encoder, followed by a deinterleaver,
followed by the inverse of the second convolutional
encoder)~\cite{PTO09,WH10b}. But we can actually do better when there is just
one quantum convolutional code that the receiver needs to decode. The receiver
can perform an online decoding unitary where he proceeds with decoding the
transmitted qubits as soon as he receives them from the channel output. We
illustrate how to do so for the encoding transformation in
(\ref{eq:FGG-Clifford-encoder-gates})---the idea is similar to our technique
from the previous section.

We first need to determine how the encoder in
(\ref{eq:FGG-Clifford-encoder-gates}) transforms the logical operators of the
code, before determining its corresponding online decoder. One can check with
computer programs~\cite{W10}\ or by hand that the encoder in
(\ref{eq:FGG-Clifford-encoder-gates}) transforms the following two unencoded
logical operators%
\[%
\begin{array}
[c]{ccc}%
I & I & X\\
I & I & Z
\end{array}
\left\vert
\begin{array}
[c]{ccc}%
I & I & I\\
I & I & I
\end{array}
\right\vert
\begin{array}
[c]{ccc}%
I & I & I\\
I & I & I
\end{array}
,
\]
to the following two encoded logical operators%
\[%
\begin{array}
[c]{ccc}%
Z & Y & I\\
Y & I & Z
\end{array}
\left\vert
\begin{array}
[c]{ccc}%
X & Z & Y\\
X & Z & Y
\end{array}
\right\vert
\begin{array}
[c]{ccc}%
I & I & I\\
I & I & I
\end{array}
.
\]
Thus, our goal now is to find an online decoder to decode both the above
logical operators and the stabilizer generators in (\ref{eq:FGG-stabilizer}).
This guarantees that the receiver decodes the information qubits properly
\cite{thesis97gottesman,WB07}\ and that he can perform measurements of the
decoded ancilla qubits whose outcomes he can subsequently feed into a decoding
algorithm for the quantum convolutional code~\cite{PTO09,WH10b}. By our same
technique as before, we can deduce that the online decoder should perform the
following transformation:%
\begin{equation}%
\begin{array}
[c]{cccc}%
I^{\otimes m} & Z & Y & I\\
I^{\otimes m} & Y & I & Z\\
I^{\otimes m} & X & X & X\\
I^{\otimes m} & Z & Z & Z\\
g_{1}^{\prime} & X & Z & Y\\
g_{2}^{\prime} & X & Z & Y\\
g_{3}^{\prime} & X & Z & Y\\
g_{4}^{\prime} & Z & Y & X
\end{array}
\rightarrow%
\begin{array}
[c]{cccc}%
I & I & I & g_{1}^{\prime}\\
I & I & I & g_{2}^{\prime}\\
I & I & I & g_{3}^{\prime}\\
I & I & I & g_{4}^{\prime}\\
I & I & X & I^{\otimes m}\\
I & I & Z & I^{\otimes m}\\
Z & I & I & I^{\otimes m}\\
I & Z & I & I^{\otimes m}%
\end{array}
.\label{eq:FGG-online-decoder}%
\end{equation}
The input-output commutation relations of the above decoder are more
complicated than those from before, but they nevertheless demand that
$g_{1}^{\prime}$, \ldots, $g_{4}^{\prime}$ should satisfy the following
commutation relations:%
\begin{align}
\left\{  g_{1}^{\prime},g_{2}^{\prime}\right\}   &  =\left[  g_{1}^{\prime
},g_{3}^{\prime}\right]  =\left\{  g_{1}^{\prime},g_{4}^{\prime}\right\}
\nonumber\\
&  =\left[  g_{2}^{\prime},g_{3}^{\prime}\right]  =\left\{  g_{2}^{\prime
},g_{4}^{\prime}\right\}  =\left\{  g_{3}^{\prime},g_{4}^{\prime}\right\}
=0.\label{eq:FGG-commutation-relations}%
\end{align}
A choice of $g_{1}^{\prime}$, \ldots, $g_{4}^{\prime}$ that suffices to
implement the above transformation is $g_{1}^{\prime}=XX$,\ $g_{2}^{\prime
}=ZX$,$\ g_{3}^{\prime}=IX$,$\ g_{4}^{\prime}=IZ$, and it is not possible for
$g_{1}^{\prime}$, \ldots, $g_{4}^{\prime}$ to satisfy the commutation
relations in (\ref{eq:FGG-commutation-relations}) with fewer than two memory
qubits. The technique for determining the minimal set of generators satisfying
the above commutation relations is exactly the same as the technique used to
determine the minimal number of ebits required by an entanglement-assisted
code~\cite{arx2008wildeOEA,PhysRevA.79.062322} (we detail this in full
generality in Ref.~\cite{HHW10a}).

An online decoder executing the transformation in (\ref{eq:FGG-online-decoder}%
) with the above choice of $g_{1}^{\prime}$, \ldots, $g_{4}^{\prime}$ is
always non-catastrophic. The line of reasoning is essentially the same as in
the previous section, though the cycles for the online decoder that we should
consider are instead of the following form:%
\[%
\begin{array}
[c]{cccc}%
h_{1} & I & I & I\\
\vdots & \vdots & \vdots & \vdots\\
h_{p} & I & I & I
\end{array}
\rightarrow%
\begin{array}
[c]{cccc}%
s_{1,1} & s_{1,2} & l_{1} & h_{2}\\
\vdots & \vdots & \vdots & \vdots\\
s_{p,1} & s_{p,2} & l_{p} & h_{1}%
\end{array}
,
\]
where $h_{1}$, \ldots, $h_{p}$ are two-qubit Pauli operators. Cycles of the
above form are relevant here because we are interested in zero-physical weight
cycles that have non-zero logical weight. Then by observing the input-output
commutation relations in (\ref{eq:FGG-online-decoder}), the Pauli operators of
the memory states in the cycle should each commute with $g_{1}^{\prime}$,
\ldots, $g_{4}^{\prime}$. Again, such cycles can only be the
zero-logical-weight self-loop at the identity memory state because the only
operator commuting with all of $g_{1}^{\prime}$, \ldots, $g_{4}^{\prime}$ is
the identity operator acting on two qubits.

\section{Encoder for a Grassl-R\"{o}tteler Code}%

\begin{figure*}[ptb]
\begin{equation}
\left[
\begin{array}
[c]{cccc}
1+D+D^{4} & 1+D+D^{2}+D^{4} & 1+D^{3}+D^{4} & 1+D^{2}+D^{3}+D^{4}
\end{array}
\right]  . \label{eq:GR-classical-code}
\end{equation}
\begin{equation}
\begin{array}
[c]{cccc}
X & X & X & X\\
Z & Z & Z & Z
\end{array}
\left\vert\begin{array}
[c]{cccc}
X & X & I & I\\
Z & Z & I & I
\end{array}
\right\vert\begin{array}
[c]{cccc}
I & X & I & X\\
I & Z & I & Z
\end{array}
\left\vert\begin{array}
[c]{cccc}
I & I & X & X\\
I & I & Z & Z
\end{array}
\right\vert\begin{array}
[c]{cccc}
X & X & X & X\\
Z & Z & Z & Z
\end{array}
\label{eq:GR-quantum-code}
\end{equation}
\end{figure*}%
We illustrate our technique on one more example of a quantum convolutional
code from Ref.~\cite{GR07} in order to demonstrate how to handle more
complicated situations. The example in the second row of Figure~1 of
Ref.~\cite{GR07} is a quantum convolutional code generated from the classical
convolutional code in (\ref{eq:GR-classical-code}) (on the next page). This
generator leads to the quantum convolutional code in (\ref{eq:GR-quantum-code}%
). So the encoding unitary should act as follows:%
\begin{equation}%
\begin{array}
[c]{ccccc}%
I^{\otimes m} & Z & I & I & I\\
I^{\otimes m} & I & Z & I & I\\
g_{1} & I & I & I & I\\
g_{2} & I & I & I & I\\
g_{3} & I & I & I & I\\
g_{4} & I & I & I & I\\
g_{5} & I & I & I & I\\
g_{6} & I & I & I & I\\
g_{7} & I & I & I & I\\
g_{8} & I & I & I & I
\end{array}
\rightarrow%
\begin{array}
[c]{ccccc}%
X & X & X & X & g_{1}\\
Z & Z & Z & Z & g_{2}\\
X & X & I & I & g_{3}\\
Z & Z & I & I & g_{4}\\
I & X & I & X & g_{5}\\
I & Z & I & Z & g_{6}\\
I & I & X & X & g_{7}\\
I & I & Z & Z & g_{8}\\
X & X & X & X & I^{\otimes m}\\
Z & Z & Z & Z & I^{\otimes m}%
\end{array}
,\label{eq:GR-transformation}%
\end{equation}
where $g_{1}$, \ldots, $g_{8}$ are Pauli operators acting on the memory
qubits. In order for a Clifford encoder preserving the input-output
commutation relations to exist, $g_{3}$ and $g_{6}$ should form an
anticommuting pair commuting with all other memory operators, $g_{4}$ and
$g_{5}$ should as well, and $g_{1}$, $g_{2}$, $g_{7}$, $g_{8}$ should commute
with each other. Thus, the encoder requires six memory qubits at minimum, and
a minimal set of generators that satisfies the commutation relations is%
\begin{align}
g_{1} &  =Z_{1},\ g_{2}=Z_{2},\ g_{3}=X_{3},\ g_{4}=X_{4},\nonumber\\
g_{5} &  =Z_{4},\ g_{6}=Z_{3},\ g_{7}=Z_{5},\ g_{8}=Z_{6}%
.\label{eq:GR-memory-choice}%
\end{align}
This number of memory qubits is in fact minimal over all
different representations of the original quantum convolutional code \cite{HHW10a}.
(Again, this task of minimizing memory is the same as minimizing the amount of
entanglement required for an entanglement-assisted
code~\cite{arx2008wildeOEA,PhysRevA.79.062322}.) Thus, an encoder implementing
the transformation in (\ref{eq:GR-transformation}) always exists by
Lemma~4\ in Appendix~B of Ref.~\cite{BFG06}.

Determining a non-catastrophic encoder for this example is a bit more
complicated than for our previous examples because the Pauli operators $g_{1}%
$, \ldots, $g_{8}$ do not form a complete basis for Pauli operators acting on
the memory qubits. Nevertheless, we can set some further constraints on the
encoder to ensure that the resulting choice is non-catastrophic. First, we
find a set of Pauli operators that completes the basis for the Pauli operators
acting on the memory. For our example, the following choice suffices:
$g_{9}=X_{1}$, $g_{10}=X_{2}$, $g_{11}=X_{5}$, $g_{12}=X_{6}$. Next, we
determine that all elements of a cycle of zero physical weight with non-zero
logical weight have the following form:%
\[%
\begin{array}
[c]{ccccc}%
h_{i} & s_{i,1} & s_{i,2} & l_{i,1} & l_{i,2}%
\end{array}
\rightarrow%
\begin{array}
[c]{ccccc}%
I & I & I & I & h_{i+1}%
\end{array}
.
\]
Thus, all memory states $h_{i}$ that are part of such a cycle should commute
with the last two output rows in (\ref{eq:GR-transformation}). They should
then commute with the last two input rows in (\ref{eq:GR-transformation}) in
order to be consistent with the input-output commutation relations of the
encoder. This restricts them to have the form $g\otimes Z^{e_{1}}\otimes
Z^{e_{2}}$ where $g$ is some four-qubit Pauli operator and $e_{1}$ and $e_{2}$
are binary numbers. We then observe that each transition $I^{\otimes4}\otimes
g\otimes Z^{e_{1}}\otimes Z^{e_{2}}$ of a cycle should commute with the
seventh and eighth rows of the output part in (\ref{eq:GR-transformation})
(under the choice of $g_{1}$, \ldots, $g_{8}$ in (\ref{eq:GR-memory-choice})),
and this further restricts the memory states that are part of such a cycle to
have the form $Z^{e_{3}}\otimes Z^{e_{4}}\otimes g^{\prime}\otimes Z^{e_{1}%
}\otimes Z^{e_{2}}$ so that they are consistent with the input-output
commutation relations (where $g^{\prime}$ is a two-qubit Pauli operator).
Continuing in this fashion, we can finally determine that states in such a
cycle should have the form $Z^{e_{3}}\otimes Z^{e_{4}}\otimes I^{\otimes
2}\otimes Z^{e_{1}}\otimes Z^{e_{2}}$. We can then eliminate cycles of this
form with non-zero logical weight by choosing extra input-output relations for
the encoder that are consistent with its input-output commutation relations
while forcing the only cycle of zero physical weight to be the self-loop at
the identity memory state. Such a choice for our example is as follows:%
\begin{multline*}%
\begin{array}
[c]{cccccccccc}%
X & I & I & I & I & I & I & I & I & I\\
I & X & I & I & I & I & I & I & I & I\\
I & I & I & I & X & I & I & I & I & I\\
I & I & I & I & I & X & I & I & I & I
\end{array}
\rightarrow\\%
\begin{array}
[c]{cccccccccc}%
I & I & I & I & I & I & Z & I & I & I\\
I & I & I & I & I & I & I & Z & I & I\\
I & Z & Z & Z & X & I & Z & I & I & I\\
X & I & I & I & I & X & I & Z & I & Z
\end{array}
\end{multline*}
Any encoder that implements the full transformation specified by the above
relations and those in (\ref{eq:GR-transformation}) is non-catastrophic. We
elaborate our technique in full generality in Ref.~\cite{HHW10a}.

\section{Simulation Results}

Our encoder in (\ref{eq:FGG-Clifford-encoder-gates})\ gives a dramatic
reduction in the quantum memory required to encode the FGG code, and this
reduction implies that the running time of a decoding algorithm for this code
becomes within reason for computer simulation---it reduces from $O\left(
4^{15}N\right)  $ to $O\left(  4N\right)  $ where $N$ is the length of the
code. We can construct a quantum turbo code with the FGG\ encoder playing the
role of both the inner and outer encoder~\cite{PTO09,WH10b}, and the running
time of the decoding algorithm is the same order in $N$ and the memory.%
\begin{figure}
[ptb]
\begin{center}
\includegraphics[
natheight=4.193500in,
natwidth=5.133500in,
width=3.5381in
]%
{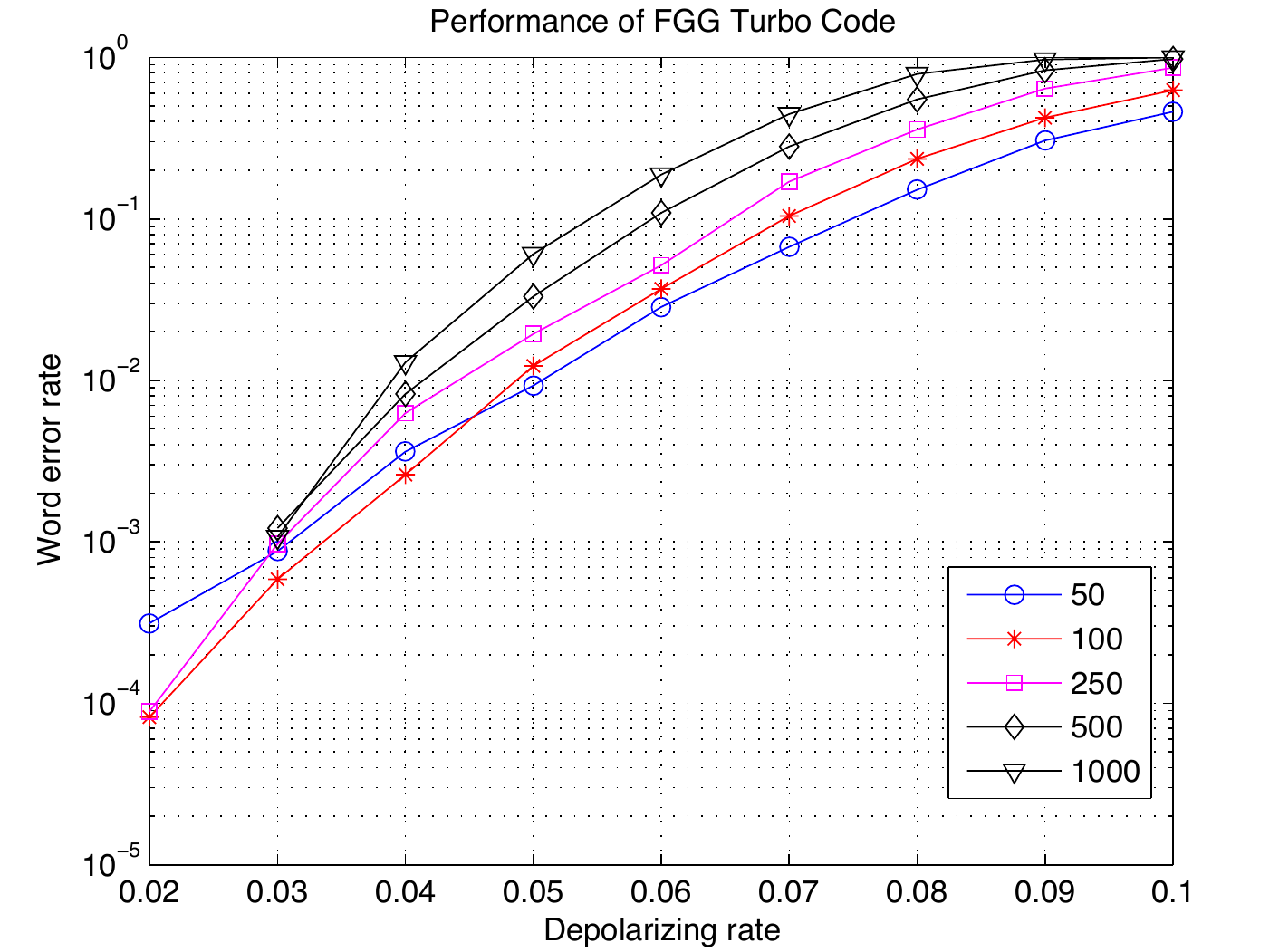}%
\caption{The performance of the rate-$1/9$ FGG\ turbo code on a depolarizing
channel. The FGG\ turbo code does not exhibit a pseudothreshold and thus has
inferior performance when compared to the codes from Refs.~\cite{PTO09,WH10b}%
.}%
\label{fig:FGG-turbo}%
\end{center}
\end{figure}

We simulated the performance of this rate-$1/9$ \textquotedblleft FGG\ turbo
code\textquotedblright\ on a depolarizing channel using free software
\cite{W10} and according to the method outlined in Section~VI of
Ref.~\cite{WH10b}. Figure~\ref{fig:FGG-turbo} demonstrates that this turbo
code unfortunately does not exhibit a pseudothreshold~\cite{PTO09,WH10b},
where performance increases as the code grows larger if the noise level is
below the pseudothreshold and it decreases with increasing code length if the
noise level is higher than the pseudothreshold. This likely has to do with the
FGG\ code's inferior distance spectrum when compared to the codes from
Refs.~\cite{PTO09,WH10b}. But the FGG\ turbo code only uses one memory qubit
as opposed to three memory qubits, and the trade-off is a decrease in
performance for a decrease in decoding time.

\section{Conclusion}

We have outlined a technique for determining minimal-memory, non-catastrophic,
polynomial-depth quantum convolutional encoders for several example codes. One
benefit of our approach is that we can circumvent the complexity issues of the
Grassl-Rotteler approach for encoding quantum convolutional codes~\cite{GR06b}.
MMW acknowledges support from the MDEIE\ (Qu\'{e}bec) PSR-SIIRI international
collaboration grant, and MH and SHK acknowledge support from the Iranian Telecommunication Research Center (ITRC). The authors thank M.-H.~Hsieh and J.~G\"{u}tschow for reading the manuscript.

\bibliographystyle{IEEEtran}
\bibliography{Ref}

\end{document}